International Journal of Computing Sciences Research (ISSN print: 2546-0552; ISSN online: 2546-115X)
Vol. 9, pp. 3602-3617
doi: 10.25147/ijcsr.2017.001.1.235
https://stepacademic.netShort Paper

# Enhancing Collaboration Through Google Workspace: Assessing and Strengthening Current Practices

Alexander B. Pahayahay
College of Computing and Information Sciences, University of Makati, Philippines
alexander.pahayahay@umak.edu.phDate received: September 5, 2024
Date received in revised form: February 27, 2025
Date accepted: March 27, 2025

Recommended citation:

Pahayahay, A. B. (2025). Enhancing collaboration through Google Workspace: Assessing and strengthening current practices. *International Journal of Computing Sciences Research, 9,* 3602-3617. https://doi.org/10.25147/ijcsr.2017.001.1.235## Abstract

*Purpose* – This study investigates the effectiveness of Google Workspace in fostering collaboration within academic settings, specifically at the University of Makati. The aim is to evaluate its role in enhancing blended learning practices and identify areas for improvement among faculty, staff, and students.

*Method* – A survey was conducted with 50 participants, including academic staff, faculty, and students at the University of Makati who regularly use Google Workspace for academic and collaborative activities. Participants were selected through purposive sampling to ensure familiarity with the platform. The study employed a quantitative research design using structured surveys to assess user experiences with key features such as real-time document editing, communication tools, etc.

*Finding* – The study found that Google Workspace and rated as "Very Effective" (mean score of 4.61) in promoting teamwork. Key advantages included improved collaboration, enhanced communication, and efficient management of group projects. However, several challenges were also noted, including low user adoption rates, limited Google Drive storage capacity, the need for better technical support, and limited offline functionality.

*Conclusion* – Google Workspace significantly supports academic collaboration in the normal practices within the University of Makati, however, it faces challenges that impact its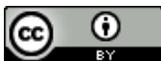

This is an Open Access article distributed under the terms of the Creative Commons Attribution License (http://creativecommons.org/licenses/by/4.0), which permits unrestricted use, distribution, and reproduction in any medium, provided the original work is properly credited.

overall effectiveness. Addressing these issues could improve user experience and platform efficiency in educational contexts.

*Recommendation* – It is recommended to enhance user adoption through targeted training and improve offline capabilities. Additionally, providing more advanced technical support could mitigate existing challenges.

*Implication* – The study contributes to the growing body of research on digital collaboration tools, offering insights into how educational institutions can optimize the use of platforms like Google Workspace for academic purposes.

## INTRODUCTION

In the digital age, remote work and distributed teams have become essential for contemporary businesses. Online collaboration solutions have become vital for fostering cooperation and communication among individuals who are in various geographical areas. Leading platform Google Workspace provides a central focus for effective collaboration with a variety of features including video conferencing, document editing, file sharing, and instant messaging. Google Documents, Google Sheets, Google Slides, and Google Meet are all part of what was once known as G Suite. Real-time collaboration on projects, spreadsheet development, data recording, and financial analysis chores are made possible by these linked tools. Creating powerful presentations is made easier with Google Slides, enabling team members to present from a distance. Effective email, chat, and video conferencing are made possible by communication platforms like Gmail and Google Meet, which improve teamwork even further. As companies carry on embracing remote work and distributed teams, Google Workspace's role in facilitating seamless collaboration is expected to become even more critical in driving organizational success. With an emphasis on its possible uses in geography education, the essay explores the usage of Google Classroom for blended learning for students studying geography.

Cartography and Basics of Topography, Population Geography, Information Systems, and Technologies in the Tourism Industry, Regional Economic and Social World Geography (Europe and the CIS), Regional Economic and Social World Geography (Africa, Latin America, Asia, Anglo-America, Australia, and Oceania), and Socio-Economic Cartography are just a few of the in-person and online courses that the authors have evaluated. The benefits of utilizing Google Classroom are highlighted in the essay, including assessing training quality, guaranteeing consistency between in-class and out-of-class learning, and real-time engagement. It also draws attention to the drawbacks of utilizing Google Classroom, including the lack of direction, low preparedness for classroom work, lack of content and technical assistance, and students' lack of internal motivation (Bondarenko et al., 2019).

## *Background of the Study*

Organizational practices and team relationships have been profoundly affected by the digital era, especially when it comes to collaboration. Businesses need to use Google Workspace, a package of cloud-based productivity tools, to establish remote and international work arrangements. This study assesses Google Workspace as an all-inclusive platform for collaboration, communication, and task management utilized by enterprises. It seeks to pinpoint possible problems and best practices within this framework for digital transformation to improve its usefulness.

In response to the growing demand for integrated technology in modern businesses, Google Workspace was developed. The goal of the study is to comprehend the context of the digital revolution of the workplace and the vital role that online collaboration tools play in the modern corporate setting. The study intends to add to the body of knowledge on digital collaboration and provide ideas for improving collaborative results by offering insights that can help firms make the most of these platforms. Even with its broad adoption, it is still necessary to evaluate its present usage patterns and pinpoint development opportunities. The COVID-19 pandemic and other worldwide events have expedited the transition towards remote work, which emphasizes the significance of dependable and effective online collaboration platforms.

Understanding a particular set of tools is simply one aspect of studying Google Workspace; another is realizing the broader picture of how the workplace is changing due to digitalization. In summary, this study intends to add to the body of knowledge on digital collaboration by providing insights and suggestions for improving Google Workspace practices. It also emphasizes the crucial role that online collaboration tools play in today's corporate environment.

## *Statement of the Problem*

This research assesses Google Workspace's collaborative practices, pointing out obstacles and suggesting ways to increase efficacy in terms of:

1. In what way does Google Workspace's user base see its success in promoting collaboration and teamwork?
2. Which Google Workspace technologies are most frequently used for collaborative activities, and what are their standard practices?
3. What difficulties arise when users utilize Google Workspace for collaboration?

4. What are the benefits can Google Workspace offer for enhancing existing collaboration practices?

## LITERATURE REVIEW

### *Computer-Supported Collaborative Systems (CSCS)*

The initiative aims to enhance learning by integrating technology into language-learning classes, preparing students for future academic and professional environments. A study conducted in a pre-university EAP course in the UAE involved thirty-one participants and explored their experiences with using Google apps for language learning assignments. Benefits highlighted by participants included online feedback, remote collaboration, and ease of use. However, students tended to divide tasks when collaborating. The study suggests that integrating new technologies into language learning can improve instruction and better equip students for future challenges (Andrew et al., 2018).

Additionally, GALL (Google Assisted Language Study) is a collaborative technology that supports both online and blended language learning. A review of 178 papers over ten years indicates that Google apps significantly enhance English language teaching, particularly writing. It also highlights the evolving role of machine translation services like Google Translate and the potential for advancements in reading, speaking, and speech recognition (Abdel-Reheem Amin, 2020).

### *Google Workspace in Education*

The COVID-19 pandemic led to a significant shift to online learning, presenting both challenges and opportunities for academic programs, especially in collaborative fields like mathematics. Initial struggles with cooperation due to personal issues and the lack of in-person interaction were overcome by students embracing digital tools for flexible and reflective learning. Combining formal and informal digital materials proved beneficial, emphasizing the importance of continuing these practices even in face-to-face settings (Selvaraj et al., 2021). The integration of technology into education has been complex, requiring ongoing adaptation and tools like Google Workspace to address challenges (Akcil et al., 2021).

Studies also explore the potential of innovative technologies like virtual and augmented reality (VR/AR) to enhance language learning, increasing motivation, engagement, and learning outcomes, and suggesting these should be more widely used in education (Huang et al., 2021). Efforts to improve digital literacy, particularly for Gen-Z students, have shown promise through interactive multimedia learning, supporting both educational and digital advancements (Natsir et al., 2022).

Post-COVID research highlights challenges in using online learning tools, including gender disparities in how faculty use these tools, but underscores the increasing importance of online platforms in educational delivery and assessment (Oguguo et al., 2023). In Georgia, studies suggested cost-effective solutions like G Suite to maintain educational continuity during the pandemic (Basilaia et al., 2020). Research in Indonesia on Google Classroom revealed its effectiveness despite limitations related to mobile access and connectivity, stressing the need to address affordability and access (Alim et al., 2019). Similarly, studies on Microsoft Teams emphasized how such platforms can bridge the digital divide in developing countries, with usability largely dependent on user experience rather than platform type (Pal & Vanijja, 2020). Collectively, these studies emphasize the importance of leveraging technology to improve educational outcomes while addressing challenges in access, usability, and engagement in a rapidly evolving educational landscape.

### *User Experience and Challenges with Google Workspace*

This study by Hafid et al. (2022) uses a qualitative research methodology to examine how SMP Negeri 1 Kedungpring manages digital classroom learning. Informants, occasions, places, and papers were some of the data sources. According to the study, preparing for digital classroom instruction with Google Workspace for Education entails organizing, arranging, carrying out, overseeing, and assessing. Creating virtual courses, establishing an account with a school domain, and utilizing Google Classroom, Google Docs, Google Sheets, Google Slides, Google Jawboard, and Google Sites are all part of the implementation. Using a Google form, the study assessed the educational process as well, and the principal, head of curriculum, and subject teachers received a report on it.

With an emphasis on the advancement of information and communication technology in education, this study attempts to comprehend how instructors use Google Workspace for Education in the context of Industry 4.0. Using a sample of ten instructors from different educational levels at SD Negeri 1 Delod Peken Tabanan, the study used a qualitative descriptive methodology. Google Meet, laptops, and stationery are among the tools used in the research for conducting interviews. With regards to simple access to information and communication technologies, such as Google Workspace for Education, which provides features like Gmail, Meet, Classroom, Drive, Docs, Slides, Forms, and Sheets, the research attempts to shed light on the needs of the modern educational period (Irani, 2022).

With Google Workspace being a leading alternative for online educators, the COVID-19 epidemic has expedited the uptake of online learning. On Malaysian architecture students' acceptance and usage of Google Workspace, however, little study has been done. Investigating how students behaved while using Google Workspace throughout the epidemic is the goal of this study. The study examined student behavior at three Universiti

Teknologi MARA (UiTM) campuses that offered design-based courses using a quantitative approach, a survey, and the Unified Theory of Acceptance and Use of Technology 2 (UTAUT2). Hedonistic Motivation had the lowest frequency distribution of behavioral patterns, according to the study, whereas Effort Expectancy had the greatest. The results imply that Google Workspace is an appropriate tool for design studio instruction and that students' adoption of using it is high (Khair et al., 2022).

### *Enhancing Collaboration with Digital Tools*

Facilitating group collaboration with instructional content in distant learning, this study suggests criteria and indicators for using digital technologies. The study of Glazunova et al. (2023) examined the capabilities available on the platforms of Google Workspace, Cisco Webex, and Microsoft 365 and assessed how well-suited they were for overseeing group projects for students in postsecondary education. The study determined that design, functional-technological, and education-communication are the three main factors to consider when choosing digital tools for group projects with educational material. The study evaluated digital tools for distant learning and established the weights assigned to these factors. Group work was found to be better served by Microsoft 365 and Google Workspace, while real-time group work worked best with Cisco Webex.

Moreover, studies such as those by Hafid et al. (2022) and Irani (2022) underscore the importance of organizing and managing digital classroom learning using Google Workspace. Both studies point to the need for proper integration and training to optimize the use of tools such as Google Docs, Google Meet, and Google Slides in educational settings.

## METHODOLOGY

Data on the utilization of Google Workspace for collaboration were analyzed using a quantitative research approach in this study. Data were gathered from a representative sample of users across various industries through an organized survey, which included closed-ended questions designed to convey comprehensive responses regarding the availability, type, and efficiency of Google Workspace features. By employing a quantitative approach, patterns, correlations, and potential causal relationships were identified through statistical analysis and objective variable measurement. The study aimed to shed light on how different user demographics and work environments influenced Google Workspace adoption and usage.

The study focused on the University of Makati's use of Google Workspace as an online collaboration tool among academic, administrative, and student staff. Focused group discussions and in-depth interviews were used to gather data on collaborative dynamics, challenges, and opportunities, to understand the use of Google Workspace products within the university context. Geographically, this research covers only within the University of Makati, ensuring a concentrated analysis of the institutional setting. The purpose of the study was to provide insights that could improve the university's use of Google Workspace and offer valuable recommendations for other educational institutions in the Philippines and beyond. The findings are expected to contribute to the broader discourse on digital collaboration in academia, emphasizing best practices and key areas for improvement in the adoption and use of online collaboration tools.

## *Population, Respondents, and Sample Size*

The academic community at the University of Makati in the Philippines, which consists of ten (10) academic staff, ten (10) faculty members, and thirty (30) students who utilize Google Workspace for academic and collaborative purposes, was the subject of this quantitative study.

## *Sampling Technique*

The study utilized the purposive sampling technique to identify participants who have an elevated level of familiarity with Google Workspace, which guaranteed the acquisition of comprehensive and insightful data. The concept of data saturation followed in determining the sample size; fifty (50) participants were involved to achieve this study, and the participants were given a questionnaire to answer. The goal of this strategic approach is to provide a thorough comprehension of collaborative activities and issues in the classroom.

## *Research Instrument and Validation*

The main study tool for evaluating Google Workspace's efficacy in cooperation is a structured questionnaire. The questionnaire asked about demographics, usage trends, difficulties encountered, and ideas for improvement. The survey evaluated respondents' opinions of the tool's influence on collaboration using a 5-point Likert scale. The frequency of utilizing different technologies for group projects is documented in the usage patterns section. To identify barriers to adopting Google Workspace for collaboration, the problems faced section included both open-ended questions and Likert scale items. Examples of sample items include complaints regarding technical difficulties and inadequate instruction on the proper use of tools. Open-ended questions were included in the section dedicated to improvement suggestions to elicit ideas from participants for improving collaborative practices. The standard for measuring reliability will be Cronbach's alpha, and content validity will be assessed by a panel of experts in digital collaboration tools.

*Statistical Treatment of Data*

This study used a structured poll with closed-ended questions to examine how Google Workspace is used in different sectors. A summary of the demographics, usage patterns, challenges faced, and recommendations for improvement of the survey participants were given. Inferential statistics were used to investigate correlations between variables and evaluate usage-related hypotheses for Google Workspace. The direction and degree of correlations between variables are evaluated using Pearson's correlation coefficient. Multiple regression research, considering variables like industrial sector, work position, and user demographics, may reveal important predictors of Google Workspace adoption and usage. Participants from various sectors have used their Google Workspace compared in groups using t-tests or ANOVA to examine variations in use. Cronbach's alpha coefficient was used to evaluate the Likert scale questions in the survey for reliability, and content validity was evaluated by a panel of experts in digital collaboration tools. Qualitative analysis identified common themes and patterns in open-ended responses. Statistical software packages like SPSS or R were used for data analysis, generating robust empirical evidence on the current state of collaborative practices using Google Workspace at the University of Makati.

## RESULTS

The evaluation of Google Workspace apps shows notable differences in how useful the various tools are viewed (Table 1). With a weighted mean of 4.77, Google Drive is the highest-rated program and is classified as "Very Effective." Its remarkable rating reflects its extensive file storage, management, and sharing options. Users frequently compliment Google Drive on its strong collaboration capabilities, ability to handle a wide range of file formats, and smooth interaction with other Google Workspace apps. With a weighted mean of 4.74, Google Meet scores well, indicating how well it supports distant collaboration and virtual meetings. Its capabilities, which include screen sharing, high-definition video quality, and Google Calendar integration, make online meetings more effective and fruitful.

The weighted averages of Google Forms and Google Docs, which are rated as "Very Effective," are 4.72 and 4.61, respectively. With a weighted mean rating of 4.02, Google Slides and Google Calendar are categorized as "Effective," indicating some areas for improvement. While Google Calendar may not fully satisfy changing expectations for sophisticated features or seamless connectivity with other productivity tools, Google Slides may provide user hurdles relating to feature constraints or integration concerns. With a weighted mean of 4.47, Google Sheets is classified as "Effective," however it falls short of the very effective tools at the top. With a weighted mean rating of 3.68, Google Chat is considered functional in that it helps with team collaboration and instant messaging; nevertheless, its total effect may be hampered by features or user experience issues. In summary, this review identifies areas where enhancements might increase functionality and customer happiness while also emphasizing the capabilities of several Google Workspace apps in providing outstanding value to users.

Table 1. Google Workspace Application

| Application | Weighted Mean | Interpretation | Rank |
|---|---|---|---|
| Gmail | 4.67 | Very Effective | 4 |
| Google Drive | 4.77 | Very Effective | 1 |
| Google Docs | 4.61 | Very Effective | 5 |
| Google Sheets | 4.47 | Effective | 6 |
| Google Slides | 4.02 | Effective | 11 |
| Calendar | 4.02 | Effective | 12 |
| Chat | 3.68 | Effective | 16 |
| Meet | 4.74 | Very Effective | 2 |
| Sites | 3.89 | Effective | 13 |
| Contacts | 3.81 | Effective | 14 |
| Classroom | 4.30 | Effective | 9 |
| Forms | 4.72 | Very Effective | 3 |
| Translate | 4.33 | Effective | 7 |
| Maps | 4.32 | Effective | 8 |
| Earth | 4.11 | Effective | 10 |
| Groups | 3.81 | Effective | 15 |

Google Workspace is rated highly in promoting teamwork and collaboration at the University of Makati, with an overall score of 4.61, indicating it is "Very Effective" (Table 2). It facilitates smooth group work, shared document editing, and real-time communication, enhancing academic collaboration. The platform's ability to solve problems is also well-rated (4.21), but there is room for improvement in response time and issue resolution. The impact of Google Workspace on collaboration is assessed at 4.56 ("Very Effective"), showing its critical role in supporting group projects. Training courses provided by Google Workspace are rated highly at 4.53 ("Very Effective"), highlighting their importance in preparing users for collaborative tasks. However, technical support, while rated effective (4.28), could be improved for faster resolution of issues.

The resistance to using Google Workspace is moderate, with a score of 4.05, suggesting some areas to address to improve user acceptance. The platform's adaptability to meet collaboration needs and its seamless integration with existing systems is well-regarded, with ratings of 4.53 and 4.49, respectively. Google Workspace also scores well in data security and privacy (4.47), ensuring users' sensitive information is protected. Finally, ease of access to collaboration tools is highly rated (4.59), reflecting the platform's user-friendly interface. Overall, Google Workspace is beneficial in enhancing teamwork, collaboration, and operational efficiency at the University of Makati, though some areas, such as technical support and user resistance, could be further improved.

Table 2. Effectivity of Google Workplace

| Indicator | Weighted Mean | Interpretation | Rank |
|---|---|---|---|
| To what extent does Google Workspace help the University of Makati promote teamwork? | 4.61 | Very Effective | 1 |
| How well does Google Workspace handle issues that arise when using it? | 4.21 | Effective | 8 |
| How do you think Google Workspace improves cooperation at Makati University? | 4.56 | Very Effective | 3 |
| How efficient is the technical assistance offered for problems relating to Google Workspace? | 4.28 | Effective | 7 |
| To what extent have the Google Workspace tool training courses helped you prepare for joint projects? | 4.53 | Very Effective | 4 |
| How do you think your team members are resisting using Google Workspace? | 4.05 | Neutral | 9 |
| To what extent can Google Workspace be tailored to meet the demands of collaboration? | 4.53 | Very Effective | 4 |
| To what extent does Google Workspace facilitate the integration of current systems and workflows? | 4.49 | Very Effective | 5 |
| To what extent do you feel comfortable with Google Workspace's data security and privacy measures? | 4.47 | Very Effective | 6 |
| To what extent does Google Workspace facilitate easy access to collaboration tools? | 4.59 | Very Effective | 2 |

## DISCUSSIONS

### *How Google Workspace Made Collaboration Easier*

Google Workspace facilitates efficient collaboration by enabling real-time document editing, virtual meetings via Google Meet, and easy file sharing through Google Drive. Its asynchronous collaboration features and centralized file storage make it easier for teams to organize and manage projects, even remotely. The platform's real-time updates, transparency, and flexibility in access allow for seamless teamwork and increased productivity.

## *Obstacles or Limitations of Google Workspace*

Challenges with Google Workspace include internet access issues, limited storage capacity (15 GB per user), and the possibility of collaboration overload, where multiple users might unintentionally edit the same sections of a document, causing confusion and inaccuracy in data output. Technical understanding may also be a barrier for some users. Privacy and data security concerns are also present, and file management can become disorganized without proper structuring. Addressing these challenges requires additional storage, better training, and secure practices.

## *Possible Areas for Improvement or Customization*

Google Workspace can be enhanced by increasing storage capacity, improving offline functionality, integrating more academic procedures, and adding specialized tools for research collaboration. Enhancements to Google Meet for large-scale meetings, along with stronger analytics and security features, would improve the platform's overall utility, particularly for academic settings. Customization options would further improve collaboration by tailoring the platform to specific needs.

## *Importance of Support and Training Programs*

The University of Makati's training initiatives are crucial for maximizing Google Workspace's potential. These programs enhance team coordination, communication, and collaboration by helping users master the platform's features. The training improves efficiency, fosters teamwork, and ensures that users can fully utilize Google Workspace's tools for project management and academic collaboration. Ongoing investment in such programs is vital for continued success.

## *Suggestions for Improving Collaboration with Google Workspace*

Proactive training and support systems can optimize Google Workspace for collaboration at the University of Makati. Integrating Google Workspace with other learning tools, addressing storage limitations through subscription plans, and enhancing Google Meet and Calendar functionalities can improve overall user experience. Focusing on data privacy, adding AI-driven features, and fostering a user forum would further enhance collaboration and efficiency at the university.

## **CONCLUSIONS**

In conclusion, Google Workspace has proven to be a highly effective tool for promoting teamwork at the University of Makati, earning a rating of 4.61. It facilitates real-

3612

time communication, collaborative document editing, and smooth group work, all of which are vital in various academic settings like remote, hybrid, and face-to-face. The high user satisfaction reflects how the platform supports teamwork, making it an essential resource for both educators and students. This positive reception aligns with the university's goals of fostering a strong academic community. The most frequently used tools for collaborative work are Google Meet, Google Drive, and Google Docs. These tools are central to the university's daily operations, enabling virtual meetings, real-time document editing, and centralized file storage.

Google Drive efficiently organizes and shares large data volumes, while Google Docs and Google Sheets allow simultaneous collaboration on documents, reducing the need for back-and-forth communication. Google Meet further facilitates virtual collaboration, ensuring that teams can work together seamlessly regardless of their location. While Google Workspace provides significant benefits, some challenges persist, including unreliable internet connections, limited storage on Google Drive, and the potential for excessive collaboration that can lead to confusion. To address these issues and improve productivity, the university would benefit from clearer policies and enhanced training on best practices. Additionally, while Google Workspace is a valuable tool for document management and remote teamwork, there are areas for improvement. Expanding storage capacity, enhancing offline functionality, and offering tailored training for varying levels of technical expertise could better meet the university's needs. Moreover, refining the platform's integration with existing workflows, such as modifying Google Meet for larger gatherings and streamlining complex group projects, could further enhance its utility at the University of Makati.

## RECOMMENDATIONS

To improve the use of Google Workspace at the University of Makati, several recommendations should be considered. First, the development and implementation of comprehensive training programs are essential to enhance the Google Workspace competency of both teachers and students. These programs should cover basic functions as well as more advanced features, such as group document editing, tool integration, and efficient data management. Regular updates about new features and best practices will help reduce technical difficulties, shorten the learning curve, and increase accuracy and efficiency in collaborative tasks. Second, the university should address the storage limitations on Google Drive to better support research and team initiatives. Exploring options to expand storage, especially for users with high data storage needs, is critical. This could involve negotiating better storage tiers with Google or developing internal guidelines for efficient data management, including regular audits and archiving systems. Third, while Google Workspace offers reliable technical support, more extensive services are needed to handle complex issues that may arise during group projects. The university should establish a specialized support team familiar with the intricacies of Google Workspace to provide prompt assistance. Incorporating automated troubleshooting tools could also minimize downtime and help address common issues quickly, ensuring smooth collaboration. Finally, the university should prioritize improving Google Workspace's offline functionality. Ensuring that all essential apps can function without connectivity issues would allow users to continue working on group projects even without internet access. Offline modifications should seamlessly sync with the cloud once connectivity is restored, ensuring continuous productivity and collaboration, regardless of location or internet stability. By addressing these areas, the university can enhance the overall experience and effectiveness of Google Workspace for academic and collaborative endeavors.

## IMPLICATIONS

The study on the effectiveness of Google Workspace at the University of Makati has several significant implications:

### *Practical Implications*

The research suggests that Google Workspace is effective in promoting teamwork and blended learning, but some areas need improvement. Key recommendations include providing targeted training for both faculty and students to improve platform usage, addressing Google Drive's storage limitations, enhancing offline functionality, and creating a dedicated support team to resolve technical challenges. These steps would improve the overall effectiveness of Google Workspace at the University of Makati, making it a more reliable tool for academic collaboration.

### *Theoretical Implications*

The study contributes to the field of digital collaboration tools by showing how platforms like Google Workspace can enhance academic collaboration. It emphasizes the importance of integrating such technology into existing educational practices and workflows to maximize its potential. The research also provides insights into which features of Google Workspace support or hinder collaboration, offering a foundation for future studies on optimizing digital tools in education.

### *Social Implications*

On a social level, the research highlights how Google Workspace can foster more inclusive and accessible learning experiences. By enabling real-time collaboration, it can bridge geographical and socio-economic gaps, promoting equity in education. The tool helps create an interconnected academic community where students, regardless of location or resources, can participate in collaborative learning. As education becomes more digital, using such tools can contribute to a more inclusive academic environment.

## ACKNOWLEDGEMENT

The author would like to express my sincere gratitude to the University of Makati community, particularly the faculty and staff of the College of Computing and Information Sciences, for their invaluable support and guidance throughout this research. My deepest appreciation goes to my family and to Raffy for their unwavering encouragement, understanding and support.

## FUNDING

The study did not receive funding from any institution.

## DECLARATIONS

### *Conflict of Interest*

The researcher declares no conflict of interest in this study.

### *Informed Consent*

Written informed consent was presented and explained the objectives and purpose to all the participants for this study.

### *Ethics Approval*

This study adheres to ethical guidelines to ensure participant integrity, privacy, and anonymity. Participants are informed about the study's goals, methods, risks, and benefits, with an emphasis on voluntary participation. Confidentiality is maintained through safe data storage and pseudonyms, while private interviews help protect privacy. Ethical approval is obtained from the appropriate review board, and researchers practice reflexivity to minimize biases. The study aims to safeguard participant rights and welfare while ensuring the reliability and validity of the results.

**Author's Biography**

Alexander B. Pahayahay previously served as the Chair of General Education for IT and as a faculty member at the University of Makati (UMak). He holds the rank of Assistant Professor II and served as an Alumni Regent at UMak, where he earned his BS in Computer Science. Additionally, he obtained a master's in information technology degree from the Technological University of the Philippines - Manila. He is actively involved in professional organizations such as the Philippine Society of Information Technology Educators, Inc. and the Computing Society of the Philippines. Currently, Alexander is pursuing a Doctorate in Information Technology at La Consolacion University of the Philippines and also serves as a part-time faculty member at Jose Rizal University, demonstrating his ongoing dedication to education and technology.